\documentclass[aps,prb,showpacs,twocolumn,superscriptaddress]{revtex4-1}
\usepackage{amsmath}
\usepackage{amssymb}
\usepackage{graphicx}
\usepackage{hyperref}
\usepackage{url}
\begin{document}
\title{Full Control of Magnetism in Manganite Bilayer by Ferroelectric Polarization}
\author{Shuai Dong}
\affiliation{Department of Physics, Southeast University, Nanjing 211189, China}
\affiliation{Department of Physics and Astronomy, University of Tennessee, Knoxville, Tennessee 37996, USA}
\affiliation{Materials Science and Technology Division, Oak Ridge National Laboratory, Oak Ridge, Tennessee 37831, USA}
\author{Elbio Dagotto}
\affiliation{Department of Physics and Astronomy, University of Tennessee, Knoxville, Tennessee 37996, USA}
\affiliation{Materials Science and Technology Division, Oak Ridge National Laboratory, Oak Ridge, Tennessee 37831, USA}
\date{\today}

\begin{abstract}
An oxide heterostructure made of manganite bilayers and ferroelectric perovskites is predicted to lead to the full control of magnetism when switching the ferroelectric polarizations. By using asymmetric polar interfaces in the superlattices, more electrons occupy the Mn layer at the $n$-type interface side than at the $p$-type side. This charge disproportionation can be enhanced or suppressed by the ferroelectric polarization. Quantum model and density functional theory calculations reach the same conclusion: a ferromagnetic-ferrimagnetic phase transition with maximal change $>90\%$ of the total magnetization can be achieved by switching the polarization's direction. This function is robust and provides full control of the magnetization's magnitude, not only its direction, via electrical methods.
\end{abstract}
\maketitle

\textit{Introduction}.
The control of magnetism using electric fields is a scientifically challenging
and technologically important subject that has attracted considerable attention
in recent years. Compared with single phase multiferroics (MFE), that typically
have a relatively poor performance, composite systems based on oxide
heterostructures involving ferroelectric (FE) (or MFE) and ferromagnetic (FM)
materials, provide more practical
alternatives.\cite{Ramesh:Nm,Vaz:Jpcm,Hwang:Nm} For example, by using
antiferromagnetic (AF) MFE layers (e.g. BiFeO$_3$, Cr$_2$O$_3$, YMnO$_3$, etc.),
the exchange bias in FM materials attached to the heterostructure can be
modulated by the FE $P$'s or
domains.\cite{Chu:Nm,Bea:Prl,Wu:Nm10,He:Nm,Laukhin:Prl,Dong:Prl2,Dong:Prb11.2}
In addition, in heterostructures magnetic anisotropies can be tuned by
electrical methods, and currents in tunneling magnetic junctions can be affected
by the FE barrier layers that also manifest as interfacial
magnetoelectricity.\cite{Shiota:Nm,Wang:Nm,Mardana:Nl,Gajek:Nm,Garcia:Sci,
Pantel:Nm}

Despite their success, from the fundamental viewpoint these controls of magnetism are relatively ``weak'' effects since the magnetic orders/moments themselves do not change substantially but only their easy axes or domain structures are rotated and tuned.

\begin{figure}
\centering
\includegraphics[width=0.5\textwidth]{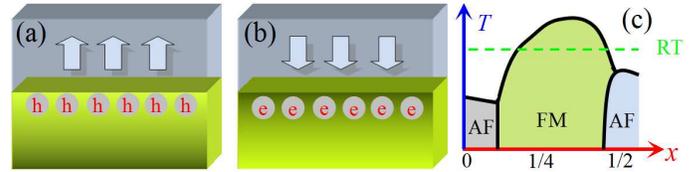}
\caption{(Color online) (a-b) Sketch of FE-field effect.
Arrows denote the FE $P$'s. Holes and electrons are attracted
to the interfaces in (a) and (b), respectively.
(c) A simplified phase diagram of LSMO.\cite{Dagotto:Prp}
$x$: doping concentration; $T$: temperature; RT: room-$T$. The FM region is sandwiched between two AF regions.}
\label{FET}
\end{figure}

Alternatively, by using the emergent properties of correlated electronic materials, more dramatic magnetoelectric (ME) effects could be envisioned in oxide heterostructures.\cite{Ahn:Rmp} For example, in FE-La$_{1-x}$Sr$_x$MnO$_3$ (LSMO) heterostructures, experiments have found giant changes in conductances triggered by switchable FE $P$'s.\cite{Vaz:Prl,Vaz:Apl,Jiang:Apl,Leufke:Prb,Yin:Nm} The associated physical mechanism is believed to be the modulation by the FE field-effect of the local electronic density in manganites near the interfaces (see Fig.~\ref{FET}).\cite{Burton:Prl,Burton:Prb,Dong:Prb11,Chen:Prb12}

However, typically this effect can only penetrate no more than $3$ unit cells (u.c.) in manganites, before the effect is almost fully screened.\cite{Burton:Prl,Burton:Prb,Dong:Prb11,Chen:Prb12} Then, the proposed magnetic phase transitions occur only within a few interfacial layers, inducing a relatively small modification of the total magnetization ($M$).\cite{Vaz:Prl,Vaz:Apl,Jiang:Apl,Leufke:Prb,Yin:Nm}

Another issue of much relevance in oxide heterostructures
is the polar discontinuity, which is emphasized for interfaces
between insulators (e.g. LaAlO$_3$-SrTiO$_3$).\cite{Nakagawa:Nm} But this effect
is often neglected in heterostructures with conductive components, e.g. LSMO,
since it will be screened within a few u.c. similarly as in the FE field-effect.

\textit{Model system}.
In this Rapid Communication, by reducing the thickness of the manganite component
to bilayer size, both the FE field effect and polar discontinuity becomes
prominent despite the metallicity of the manganite. A direct advantage of bilayers
is the maximized interface/volume ratio (up to $100\%$) for manganites that allows
each manganite layer to be fully controlled by the FE $P$. More importantly,
the neighboring asymmetric polar interfaces break the symmetry of the FE field-effect
in periodic SLs, conceptually different from results in symmetric interfaces in SLs
or single interfaces in simple heterostructures. The asymmetric design and ultra-thin
bilayers are crucial to achieve the full control of magnetism reported in our study.

\begin{figure}
\centering
\includegraphics[width=0.47\textwidth]{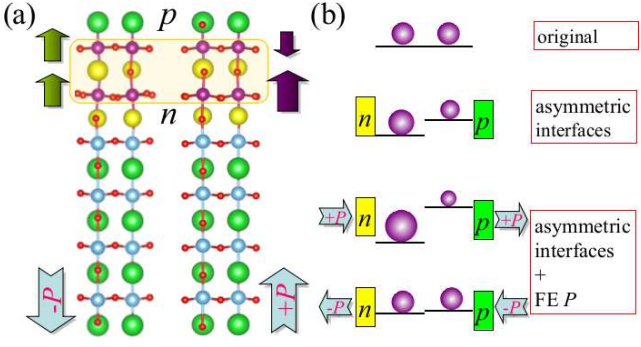}
\caption{(Color online) (a) Sketch of crystal structure.\cite{Momma:Jac} Green=$D$; red=O; cyan=Ti; purple=Mn; yellow=$R_{1-x}A_x$. The $n$-/$p$-type interfaces are indicated. Left/right are the $-P$/$+P$ cases, with switched magnetic orders (FM/AF).  (b) The $e_{\rm g}$ density (spheres) and potential (bars) modulated by asymmetric interfaces (bricks) and FE $P$ (arrows). }
\label{crystal}
\end{figure}

As our model system, SLs stacked along the conventional (001)-direction made of
$R_{1-x}A_x$MnO$_3$-$D$TiO$_3$ ($D$=divalent cation, $R$=trivalent rare-earth,
and $A$=divalent alkaline-earth) are here considered, as sketched in Fig.~\ref{crystal}(a).
The manganite components are thin involving only bilayers while the FE titanate is assumed to be slightly thicker to maintain its $P$.\cite{Jiang:Apl} Asymmetric polar interfaces are used: the interfaces with TiO$_2$-$R_{1-x}A_x$O-MnO$_2$ and TiO$_2$-$D$O-MnO$_2$ will be referred to as $n$-type and $p$-type interfaces, respectively. The $n$-type interface, with a positively charged ($R_{1-x}A_x$O)$^{(1-x)+}$ layer, will attract electrons to its nearest-neighbor (NN) MnO$_2$ layer, while the $p$-type interface will repel electrons away from the interface. Therefore, even without ferroelectricity the asymmetric interfaces already modulate the electronic density and electrostatic potential within the manganite bilayers.

When the FE $P$ points to the $n$-type interface (the $+P$ case), the electrostatic potential difference between the two MnO$_2$ layers will be further split, thus enhancing the charge disproportionation. However, when the FE $P$ points to the $p$-type interface (the $-P$ case) the electrostatic potential from the polar interfaces will be partially compensated, thus suppressing the electronic disproportionation. The above processes are summarized in Fig.~\ref{crystal}(b). In the ideal limit of $-P$ case, if these two effects (asymmetric polar interfaces vs. FE $P$) could be fully balanced, both the electrostatic potential and electronic distribution in the manganite bilayers would become uniform. By suitable combinations of couplings, this nearly full compensation is possible since a robust FE perovskite (e.g. PbZr$_y$Ti$_{1-y}$O$_3$ or Ba$_{1-y}$Sr$_y$TiO$_3$) has a large $P$, which is equivalent to a surface charge of $0.1-1$ electrons per u.c.
that can be tuned by adjusting the concentration $y$ to fit the polar charge ($R_{1-x}A_x$O)$^{(1-x)+}$ which is very similar in magnitude. Below, this ideal $-P$ case limit is adopted in the model simulations (with FE surface charge $(1-x)/2$ electrons per u.c.) to achieve a clear physical scenario and magnify contrasting effects when compared with the $+P$ case. Deviations from this ideal limit lead to qualitatively similar results in practice.

\textit{Methods}.
The two-orbital double-exchange model with both the NN superexchange and electron-lattice coupling is here employed for the manganite bilayer components.~\cite{Dagotto:Prp} The effects of the FE $P$ and polar layers are modeled by an electrostatic potential.\cite{Dong:Prb11} A $6$$\times$$6$$\times$$2$ cluster is used to simulate the manganite bilayer. In-plane ``twisted'' boundary conditions (BC) are adopted in the zero-temperature ($T$) self-consistent calculations to reduce finite-size effects,\cite{Dong:Prb12} while periodic BC are used in the computer time-consuming finite-$T$ Monte Carlo (MC) simulations. The average
$e_{\rm g}$ electronic density ($\langle n \rangle$) in the manganite bilayer is chosen as $0.7083$, corresponding to a regime that typically has a FM ground state in manganites. All energies will be in units of $t_0$, the double-exchange hopping amplitude ($\approx0.4-0.5$ eV for LSMO).\cite{Dagotto:Prp,Dong:Prb11} In addition, DFT calculations were performed on the BaTiO$_3$-LSMO SL using the Vienna \textit{ab initio} Simulation Package (VASP).\cite{Kresse:Prb,Kresse:Prb96} Details of model Hamiltonian and numerical methods are in the supplementary material.\cite{Supp}

\begin{figure}
\centering
\includegraphics[width=0.5\textwidth]{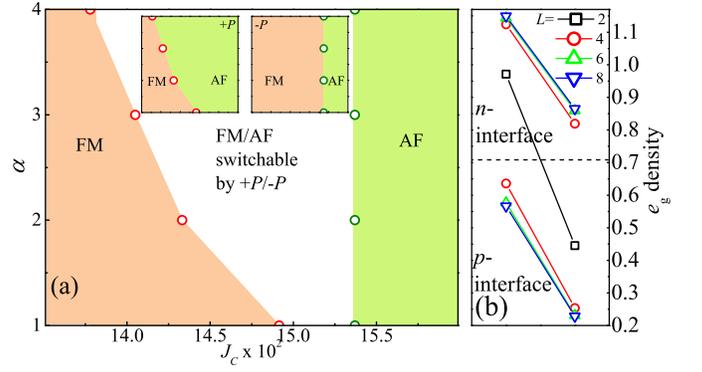}
\caption{(Color online) Zero-$T$ model simulation results. (a) The ground state phase diagram of manganite bilayers
under $\pm P$.
The middle (white) region is magnetically switchable by the FE $P$. Insets: phase diagrams under $+P$ (left) and $-P$ (right), respectively.
The dielectric constant $\varepsilon$ is represented by a Coulombic coefficient $\alpha\sim1/\varepsilon$ ($\alpha=1$ corresponds to $\varepsilon=90$).\cite{Supp} (b) The $+P$ modulated electronic densities of two-interfacial-layers with various manganite lengths ($L$). Dashed line: the original density. For all large $L$ ($>2$) cases, the $n$-type interfaces own ``high+high" density profiles while they are ``low+low" for the $p$-type ones, which can resemble phase transitions in bulks. Only for the $L=2$ case, the density profile is ``high+low", which can not be mapped to the bulk's phase diagram directly.}
\label{ZT}
\end{figure}

\textit{Zero-$T$ self-consistent calculation}.
First, the electrostatic potentials affecting the $e_{\rm g}$ electrons and the associated $e_{\rm g}$ densities ($n_i$: $i$ is the layer index) are calculated at $T=0$ self-consistently via the Poisson equation and the model Hamiltonian. Then, the energies of the FM and AF states
are compared to determine the ground state under $\pm P$. As shown in Fig.~\ref{ZT}(a), in the $-P$ case the ground state is FM if the interlayer superexchange coupling $J_c$ is smaller than $0.153$. In the $+P$ case, the ground state is AF for a $J_c$ larger than specific values that depend on the dielectric constant ($\varepsilon$) of manganite bilayers, all lower than $0.153$ for all $\varepsilon$'s studied here. Therefore, within the middle region the magnetic ground state can switch from FM to AF by switching the direction of the FE $P$. Strictly speaking, here the AF order is ferrimagnetic once the magnetic moments from the $e_{\rm g}$ electrons are taken into account, since $n_1>n_2$. Thus, the magnetic switch occurs between a FM state with strong $M_{\rm FM}$ and an AF state with a much weaker $M_{\rm AF}$. Even with this caveat, the variation in $M$ created by the $P$ switch remains quite significant: $1-M_{\rm AF}/M_{\rm FM}=1-[(3+n_1)-(3+n_2)]/[(3+n_1)+(3+n_2)]=92\%$ ideally.

It is interesting to compare the magnetic switch effects described here against the interfacial phase transitions studied in thick manganite-FE heterostructures.\cite{Burton:Prl,Burton:Prb,Dong:Prb11,Chen:Prb12} As shown in Fig.~\ref{ZT}(b), in thick manganite layers the local electronic densities of the first two-interfacial-layers can both be substantially enhanced or suppressed by the field effect. Then,
the exchange coupling between the first two-interfacial layers can be intuitively guessed from the bulk's phase diagram. For example,
if the local densities of both layers are close to $1$, it is natural to expect locally an A-type AF state.\cite{Dagotto:Prp} By contrast, this expectation is unrealistic in the bilayer case, since once $n_1$ is close to $1$ then $n_2$ must be close or below $0.5$. In this sense, the FE field-effect in the bilayers is anomalous, with strong interference effects between the $n$-/$p$- interfaces. And the magnetic coupling between the $n_1\approx1$ and $n_2\approx0.5$ layers is
unclear \textit{a priori}. Thus, in spite of similarities, the underlying mechanism of magnetic switch in the bilayers studied here is not qualitatively the same as for the phase transitions reported in thicker cases. Furthermore, due to the compensation effect between the neighboring $n$- and $p$-type interfaces, the charge modulation and electrostatic potential between the top and bottom layers in bilayers are obviously weaker than for thicker cases (more details in supplementary materials).\cite{Supp}

\begin{figure}
\centering
\includegraphics[width=0.33\textwidth]{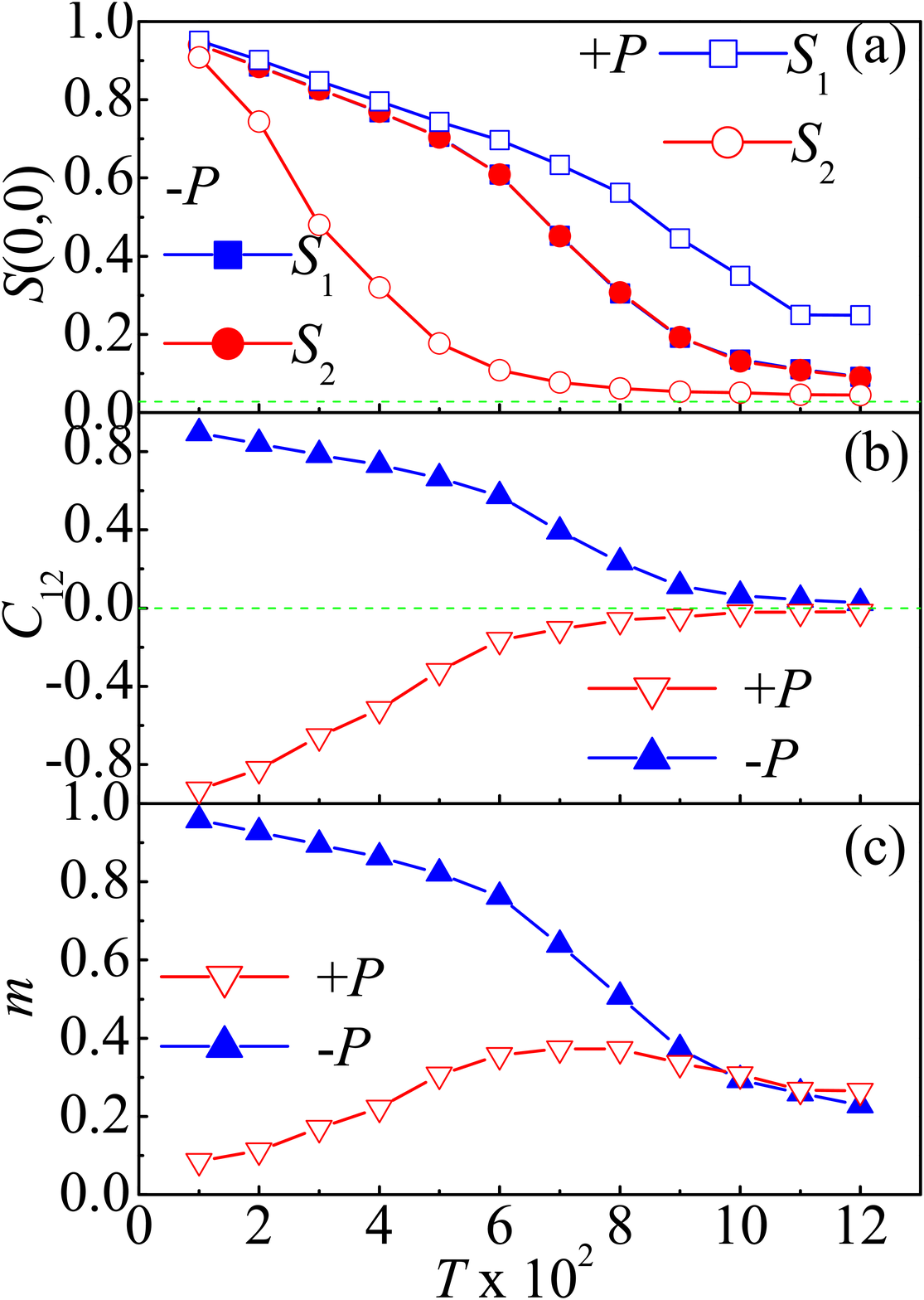}\includegraphics[width=0.15\textwidth]{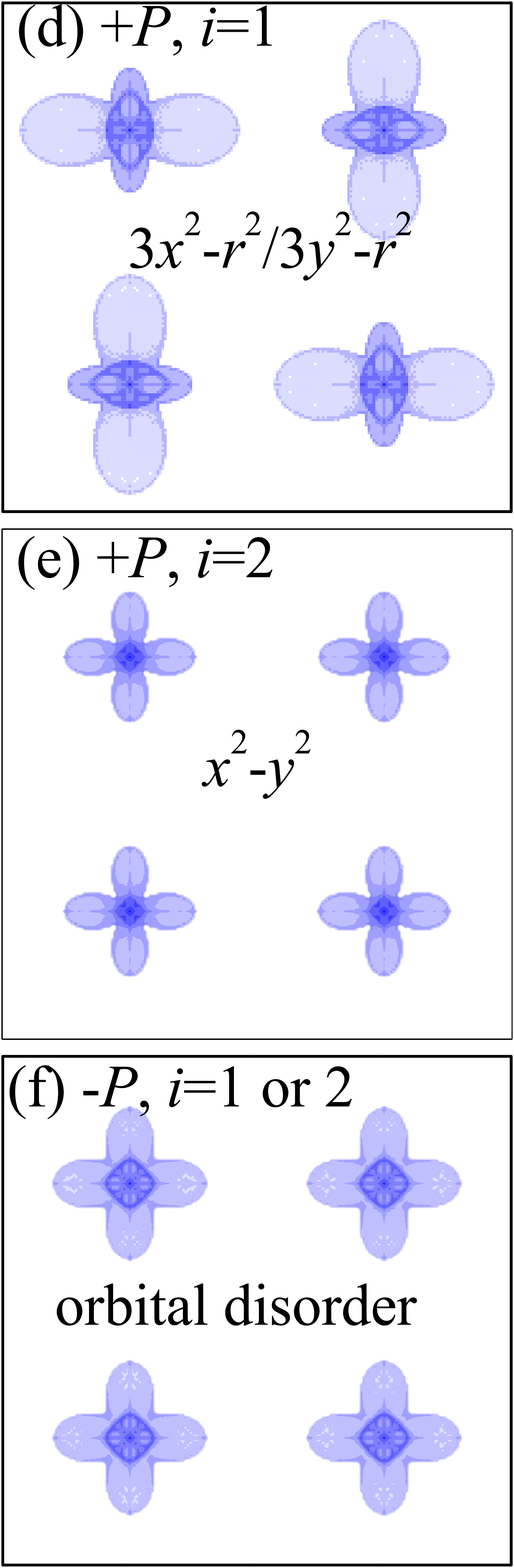}
\caption{(Color online) Results of MC simulations varying $T$ for both $\pm P$.
(a) Normalized spin structure factor at wavevector ($0$,$0$) (corresponding to
a ``FM'' order) for each layer. (b) Average NN spin correlation ($C_{12}$)
between two layers. (c) Normalized $M$ ($m$) of the $t_{\rm 2g}$ spin textures.
(d-f) Orbitals patterns (top view) obtained from MC simulations at $T=0.01$. The size of
the orbital lobes is proportional to the local electronic density. }
\label{MC}
\end{figure}

\textit{Finite-T MC simulation}.
The calculations described thus far relied on the comparison of energies between the ideal FM and AF phases, which is intuitive correct but needs confirmation using more powerful many-body techniques. In the following, finite-$T$ MC simulations will be employed to confirm the previous results. The electron-phonon coupling is also taken into account in the MC simulation. Here the electron-phonon coupling coefficient is chosen as $1.2$ and the superexchange coefficients as $J_{ab}=0.07$ in-plane and $J_c=0.14$ out-of-plane, which are realistic values from previous studies of manganites.\cite{Dagotto:Prp,Dong:Prb11}

In this more rigorous approach, the switch from FM to AF states is still observed. Fig.~\ref{MC}(a) shows the layer-resolved in-plane spin structure factors at wavevector ($0$,$0$) corresponding to in-plane ``FM" order. All curves show paramagnetic (PM) to ``FM" transitions although at different $T_{\rm C}$'s. In the $-P$ case, the two layers transit synchronically since the electronic density is uniform. By contrast, in the $+P$ case, the two layers transit separately. Thus, in the mid-$T$ region, the $1$st layer becomes ``FM", but the $2$nd layer remains PM.

The averaged NN spin correlations between layers are shown in Fig.~\ref{MC}(b). Clearly, with decreasing $T$ the $+P$ and $-P$ cases show opposite tendencies (AF vs. FM coupling). Thus, the transitions revealed in Fig.~\ref{MC}(a) are PM-FM for the $-P$ case but PM-AF (ferrimagnetic) for the $+P$ case. The average $M$'s of the $t_{\rm 2g}$ textures are presented in Fig.~\ref{MC}(c), which also display a clear contrast. In the $-P$ case, there is a peak in $M$ in the mid-$T$ region suggesting a ferrimagnetic transition, with a small but nonzero $M$ up to low temperatures.

The $T_{\rm C}$'s observed in these MC simulations are quite high. For a rough estimation, if the $T_{\rm C}$ of the bilayer ($\approx0.09$) in the $-P$ case is used to fit the $T_{\rm C}$ ($\approx375$ K) corresponding to LSMO ($x\approx0.3$),\cite{Dagotto:Prp} the upper-limit working $T$ ($\approx0.07-0.08$) for the magnetic switch can reach up to $290-330$ K. Also, if the energy unit $t_0$ is estimated to be $\approx0.4-0.5$ eV for LSMO,\cite{Dagotto:Prp} the upper-limit working $T$ grows to $310-380$ K. Both these estimations suggest that our proposed setup works at room-$T$. Of course, finite-size extrapolations are difficult via time-consuming MC techniques.

Compared with the $T=0$ self-consistent calculation using two preset candidate phases, the unbiased finite-$T$ MC simulation is more reliable. For example, in our MC simulation, the robust electron-phonon coupling is essential to stabilize the AF phase while it is not required in the $T=0$ self-consistent case. In bulk undoped manganites such as LaMnO$_3$, the staggered $3x^2-r^2$/$3y^2-r^2$ orbital order (OO) associated with the Jahn-Teller distortion is prominent and crucial for the A-type AF state.\cite{Dagotto:Prp,Hotta:Rpp} This OO also plays an important role here in the bilayers. As shown in Fig.~\ref{MC}(d-e), the $+P$ case displays different OO for the two layers. The first layer, which is close to the undoped case, shows the $3x^2-r^2$/$3y^2-r^2$-like OO (Fig.~\ref{MC}(d)). However, in the second layer, with an average electronic density slightly below $0.5$, another type of OO is present which agrees with the $x^2-y^2$-type OO  (Fig.~\ref{MC}(e)) known to exist in half-doped manganites.\cite{Dagotto:Prp,Hotta:Rpp} Both these two OO's have strong orbital lobes lying in-plane, which enhances (suppresses) the in-plane (out-of-plane) double-exchange processes. Thus, this hybrid OO's combination is advantageous to stabilize the AF order in such a bilayer. By contrast, the $-P$ case shows uniform orbital occupancy (Fig.~\ref{MC}(f)), which prefers the FM state.

\textit{DFT study}.
The model simulations described above have been carried out in the ideal $-P$ limit (i.e. with full compensation between FE $P$ and polar interfaces). However, as stated before, our predictions are not restricted by this condition. To confirm the robustness of our proposal, a preliminary \textit{ab-initio} DFT calculation was performed to verify the FE control of magnetic order.

\begin{table}
\caption{DFT results. The first two columns specify the initial conditions. The energy differences (per Mn) between the FM (reference state) and AF orders are in meV units. $m_1$ and $m_2$ denote the local magnetic moment for the Mn cations using Wigner-Seitz spheres as specified by VASP which is not accurate but qualitative preferable. $M$ is the net magnetization. All moments in $\mu_B$/Mn units. The modulation of local magnetic moment is almost identical to the modulation of local electron density due to the half metal character of LSMO.}
\begin{tabular*}{0.48\textwidth}{@{\extracolsep{\fill}}llllllr}
\hline \hline
FE & Order & Energy & $m_1$ & ~$m_2$ & ~$M$\\
\hline
$+P$ & FM~ & ~~~$0$ & $3.485$ & ~~$3.116$ & ~~$3.692$ \\
$+P$ & AF & $-13.15$ & $3.421$ & $-3.080$ & ~~$0.227$\\
\hline
$-P$ & FM~ & ~~~$0$ & $3.274$ & ~~$3.577$ & ~~$3.752$\\
$-P$ & AF & ~~$21.43$ & $3.144$ & $-3.571$ & $-0.242$\\
\hline \hline
\end{tabular*}
\vskip -0.45 cm
\label{dft}
\end{table}

The (BaTiO$_3$)$_4$-(LSMO)$_2$ ($x=1/4$) SL was studied as the model system, as sketched in Fig.~\ref{crystal}(a). According to the model study described above, an anisotropic superexchange is necessary for a switch function. In the DFT study, an in-plane tensile strain (for manganite) can induce such an effect. Thus, here the in-plane lattice constant of the SL is fixed to be $3.989$ \AA{} to fit the KTaO$_3$ substrate,\footnote{Some other combinations of substrates/manganites/titantes have also been tested. A proper in-plane tensile is important to tune the subtle balance between the FM and AF orders, namely the switching function.} which can provide tensile strain to the LSMO bilayer.
As shown in Table I, the calculated energies indicate that the ground state is FM under the $-P$ condition, but it switches to the AF state by using $+P$. The
local magnetic moments also show a significant modulation in magnitude, implying the cross impact of the FE $P$ and polar interfaces combination. The nonzero net $M$ of the AF state in the DFT study suggests a ferrimagnetic state. In spite of this caveat, the FM state displays a much larger net $M$, giving rise to a $93.9\%$ modulation by switching $P$, in agreement with the model calculations described above.
Then, the DFT study also confirms the FE control of magnetism, despite the
modifications of the $e_{\rm g}$ density and the use of a non-ideal $-P$ condition. More details of our DFT study can be found in the supplementary material.\cite{Supp}

\textit{Note}.
Finally, it is important to remark that although the notorious ``dead layers" problem in real ultra-thin manganite films may suppress ferromagnetism significantly,\cite{Huijben:Prb,Tebano:Prl} recent experiments indicate that the ``dead layers" of LSMO should be thinner than $2$ u.c. per interface in the SL geometries.\cite{Kourkoutis:Pnas,Gray:Prb} The latest experiments and theoretical simulations also show that local non-stoichiometry is responsible to ``dead layers".\cite{Peng:Arx,Song:Sia,Wang:Apl13} Thus, with further
improvements in the fabrication techniques ``alive" manganite bilayers in SLs will be possible,
as designed in our model. Recent experimental and theoretical progress in the ``dead layers" issue is shown in the supplementary material.\cite{Supp}.

In summary, our theoretical studies, using both models and \textit{ab-initio} methods, predict
the full control of magnetism when manganite bilayers are coupled to FE polarizations.
The combination of FE polarization and asymmetric polar interfaces gives rise to two
competing magnetic states: ferromagnetic and ferrimagnetic ones. The change of the
total magnetization is remarkable (up to $\sim90\%$) and may persist to room temperatures.
Although our study uses titanates as the FE layers, the physical mechanism is general and applicable to other ferroelectrics. Similar effects are expected when using manganites at other doping concentrations and with other bandwidths, increasing the range of compounds where our proposal can be realized. Therefore, our work provides a potential design to pursue the full control of magnetism in oxide heterostructures.

S.D. was supported by the 973 Projects of China (2011CB922101), NSFC (11004027, 11274060), NCET, and RFDP. E.D. was supported by the U.S. DOE, Office of Basic Energy Sciences, Materials Sciences and Engineering Division.


\newpage
\section{Details of model simulations}
\subsection{Model Hamiltonian and parameters}
According to Refs.~\onlinecite{Dagotto:Prp,Dagotto:Bok}, the Hamiltonian of
the
two-orbital double-exchange model reads as:
\begin{eqnarray}
\nonumber H&=&-\sum_{<ij>}^{\alpha\beta}t^{\vec{r}}_{\alpha\beta}(\Omega_{ij}c_{i\alpha}^{\dagger}c_{j\beta}+H.c.)+\sum_{<ij>}J_{\vec{r}}\vec{S}_{i}\cdot \vec{S}_{j}\\
&&+\lambda\sum_{i}(-\beta Q_{1i}n_i+Q_{2i}\tau_{xi}+Q_{3i}\tau_{zi})+\sum_{i}V_in_i.
\end{eqnarray}

The first term denotes the standard double-exchange hopping process for the
$e_{\rm g}$ electrons between nearest-neighbor sites $i$ and $j$. The operators
$c_{i\alpha}$ ($c_{i\alpha}^{\dag}$) annihilate (create) an $e_{\rm g}$ electron
at the orbital $\alpha$ of the lattice site $i$.
Within the standard infinite Hund coupling approximation,
the spin of the $e_{\rm g}$ electrons is always parallel to the
spin of the localized $t_{\rm 2g}$ degrees of freedom $\vec{S}_{i}$,
generating the Berry phase $\Omega_{ij}$ as
$\cos(\theta_{i}/2)\cos(\theta_{j}/2)+\sin(\theta_{i}/2)\sin(\theta_{j}/2)\exp[-i(\phi_{i}-\phi_{j})]$, where $\theta$ and $\phi$ are the polar and azimuthal angles
of the $t_{\rm 2g}$ spins, respectively.
The three nearest-neighbor (NN) hopping directions are denoted by $\vec{r}$.
Two $e_{\rm g}$ orbitals ($a$: $x^2-y^2$ and $b$: $3z^2-r^2$)
are involved in the double-exchange process for manganites, with the
hopping amplitudes given by:
\begin{eqnarray}
\nonumber t^x&=&\left(
\begin{array}{cc}
t^x_{aa} &  t^x_{ab} \\
t^x_{ba} &  t^x_{bb}
\end{array}
\right) =\frac{t_0}{4}\left(
\begin{array}{cc}
3 &  -\sqrt{3} \\
-\sqrt{3} &  1
\end{array}
\right),\\
\nonumber t^y&=&\left(
\begin{array}{cc}
t^y_{aa} &  t^y_{ab} \\
t^y_{ba} &  t^y_{bb}
\end{array}
\right) =\frac{t_0}{4}\left(
\begin{array}{cc}
3 &  \sqrt{3} \\
\sqrt{3} &  1
\end{array}
\right),\\
t^z&=&\left(
\begin{array}{cc}
t^z_{aa} &  t^z_{ab} \\
t^z_{ba} &  t^z_{bb}
\end{array}
\right) =t_0\left(
\begin{array}{cc}
0 &  0 \\
0 &  1
\end{array}
\right).
\end{eqnarray}
The hopping $t_0$ will be considered as the unit of energy. This hopping can be roughly estimated to be
 $\sim 0.4-0.5$ eV.\cite{Dagotto:Prp,Dagotto:Bok}

The second term of the Hamiltonian is the antiferromagnetic superexchange interaction
between the NN $t_{\rm 2g}$ spins. The typical value of the superexchange coupling
is in the order of $0.1t_0$ based on a variety of previous investigations
for bulk manganites.\cite{Dagotto:Prp,Dagotto:Bok}
In the Monte Carlo simulation described in the main text, an in-plane isotropic $J_{\vec{r}}$
is adopted since the bilayer growing along the (001) direction has a tetragonal-like
symmetry. However, the exchange along $c$-axis is different from those in-plane
especially when strain from the substrate is present.

The third term stands for the electron-lattice interaction, with $\lambda$ being
a dimensionless coupling. Both the breathing mode ($Q_1\sim[\delta_x+\delta_y+\delta_z]$)
and two Jahn-Teller modes ($Q_2\sim[\delta_x-\delta_y]$ and $Q_3\sim[2\delta_z-\delta_x-\delta_y]$) are considered here. $\delta_{r}$ stands for the
change in the length of the O-Mn-O bonds along a particular axis $r$.
$n_i$ is the local $e_{\rm g}$ electronic density.
The $\tau_{x}$ ($=c_{a}^{\dag}c_{b}+c_{b}^{\dag}c_{a}$) and $\tau_{z}$ ($=c_{a}^{\dag}c_{a}-c_{b}^{\dag}c_{b}$) are orbital pseudospin operators.

In the last term,  $V_i$ is the on-site electrostatic potential, which is layer-dependent.
In the $T$=0 self-consistent calculation, $V_i$ is determined via the one-dimensional
Poisson equation.
In the Poisson equation, $\alpha$ is used as the Coulomb coefficient. $\alpha$ is
inversely proportional
to the dielectric constant $\varepsilon$ [$\alpha=d/(\varepsilon t_0$),
where $d$ is the lattice constant and $\varepsilon$ is the dielectric constant].
When calculating the total energy, the Coulombic potential affecting
the A-site cations and ferroelectric polarization will also been added.
Readers can find more details regarding the electrostatic potential and energies
in one of our previous publications.\cite{Dong:Prb11}

A $6\times6\times2$ cluster is used to simulate the manganite bilayer. In-plane
¡°twisted¡± boundary conditions are adopted in the
zero-temperature self-consistent calculations to reduce finite-size effects.\cite{Dong:Prb11} while periodic boundary conditions are used in
the computer time-consuming finite-temperature Monte Carlo simulations.

\subsection{Zero-temperature simulations}
For a fixed set of
parameters, the electronic density and potential are calculated self-consistently at zero-temperature.
Here both the FM and AF backgrounds are considered and several values of the manganite
dielectric constant $\varepsilon$ were tested, which are characterized by a Coulombic coefficient $\alpha$
$\sim 1/\varepsilon$ ($\alpha=1$ corresponds to $\varepsilon=90$).\cite{Dong:Prb11}

\begin{figure}
\centering
\includegraphics[width=0.5\textwidth]{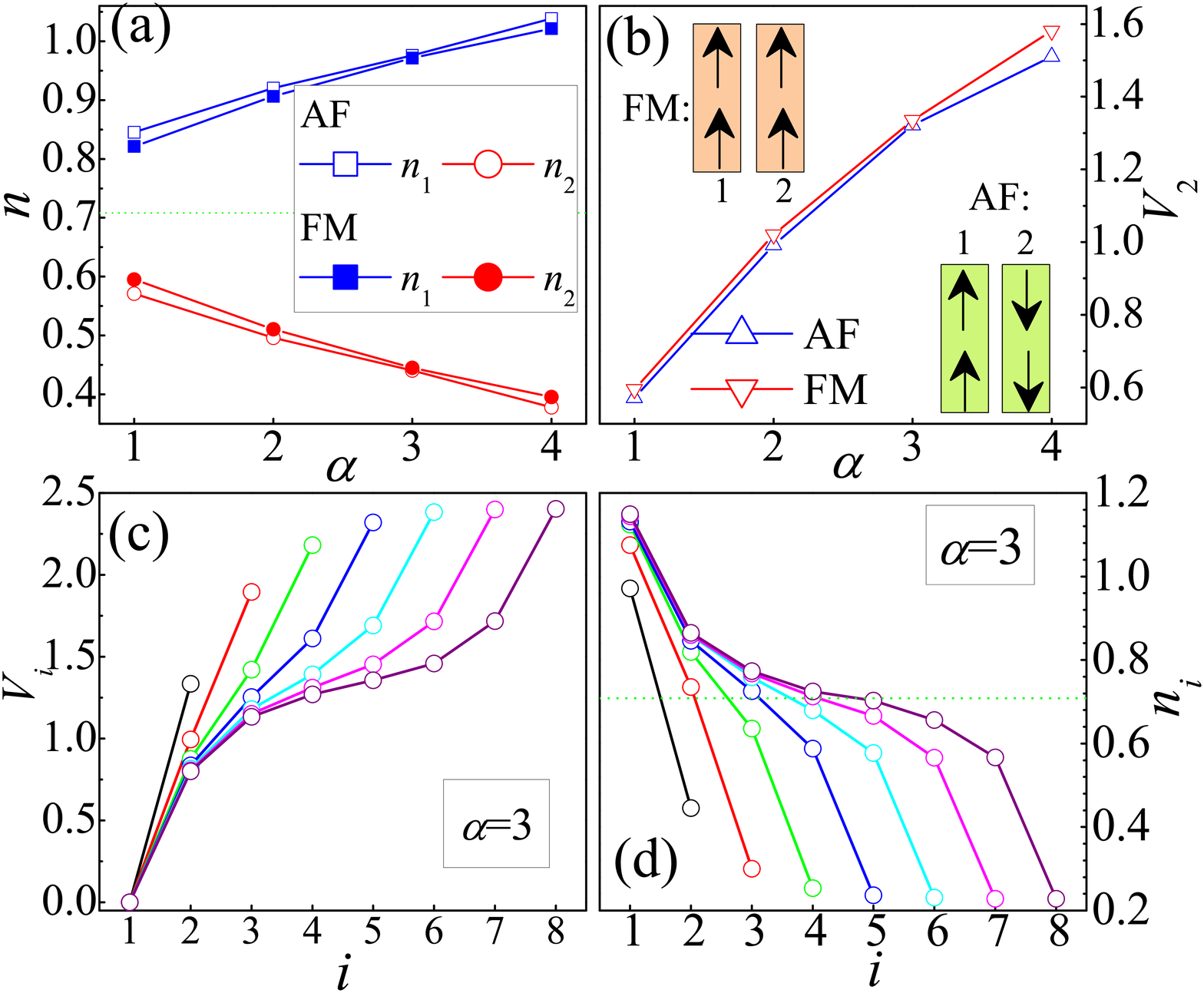}
\caption{ Results of the zero temperature self-consistent calculations under $+P$.
(a) Electronic density $n_i$ vs. Coulombic coefficient $\alpha$.
The average $e_{\rm g}$ electron density is $0.7083$ per Mn (green dot lines in (a)).
(b) Electrostatic potential difference between the two layers vs.~$\alpha$.
In (a) and (b), two sets of $t_{\rm 2g}$ spins [FM and AF, sketched as insets in (b)]
are adopted, which do not show a substantial difference.
(c) Potential profiles and (d) density profiles for various length manganite layers ($2\leq L\leq8$) when $\alpha=3$.}.
\label{zt}
\end{figure}

As shown in Fig.~\ref{zt}(a), in the $+P$ case the electronic densities
are split between the two layers with the higher value near the $n$-type interface ($i$=$1$).
Increasing the Coulombic coefficient $\alpha$ increases further this splitting,
inducing also a large potential modulation
[Fig.~\ref{zt}(b)].\cite{Dong:Prb11}
FM and AF $t_{\rm 2g}$ spin orders
[insets of Fig.~\ref{zt}(b)] are here adopted for comparison, and both
give quite similar $n_i$ and $V_i$ with only small differences.
Therefore, as a first-order approximation the electrostatic
modulation from the FE $P$ and polar interfaces will be assumed
to be independent of the manganite bilayer spin order. For
the $-P$ case, $V_i$ and $n_i$ are uniform (not shown) since
the ideal limit is used.

By using the same process, other superlattices with thicker manganite layers can also be calculated, as shown in Fig.~\ref{zt}(c-d).
It is clear that the bilayer one has the weakest charge disproportion and potential modulation, because the electrostatic potential is approximately in proportional to the thickness of manganite layers.
This weakest charge disproportion will also make the phase transition in bilayers not as easy to occur
as in the thicker layers studied before.\cite{Dong:Prb11} Fine tuning of orbital order and anisotropic exchange, as carried out in our work, are essential in the bilayer case but not required in thicker cases.

\subsection{Finite-temperature Monte Carlo simulations}
The Monte Carlo simulation is applied to the classical spin variables $\vec{S}_i$ and
the lattice distortions $\delta_{r}$ ($r=x$ or $y$, while $\delta_{z}$
is frozen at the original $0$ value), while exact diagonalization is used for
the fermionic sector (i.e. for the $e_{\rm g}$ electrons).
The first $1\times10^4$ Monte Carlo steps are adopted for thermal equilibrium and the following $1\times10^4$ Monte Carlo steps are used for measurements.
Readers can find more details of the Monte Carlo method employed
in the present work in Refs.~\onlinecite{Dagotto:Prp,Dagotto:Bok}

\begin{figure}
\centering
\includegraphics[width=0.4\textwidth]{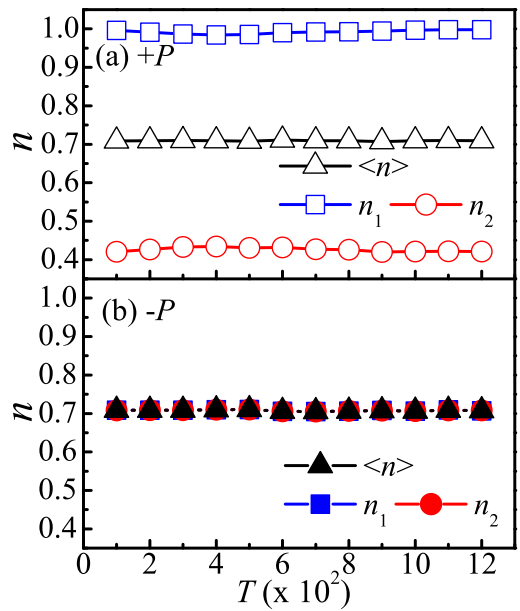}
\caption{Layer resolved electron densities obtained in Monte Carlo simulations by varying temperature for (a) $+P$ and (b) $-P$. $<n>=(n_1+n_2)/2$ is
the average density.}
\label{mc}
\end{figure}

During the Monte Carlo simulation, the electrostatic potential is fixed as a proper value $V_2-V_1=1.2$ considering the values shown in Fig.~\ref{zt}(b). This simplification depends on two preconditions: the electrostatic potential must be not affected too much by 1) the magnetic state and 2) temperature. The first one has been partially confirmed already, as shown in Fig.~\ref{zt}(a), and will be further reconfirmed
below, as shown in Fig.~\ref{mc}(a). By using FE materials with high Curie temperatures, then the second one will also be not difficult to be satisfied.

As shown in Fig.~\ref{mc}(a), the $+P$ case gives a robust charge disproportion which is almost temperature-independent in the studied range despite the magnetic-magnetic disorder-ordering transition. This result is self-consistent with the fixed electrostatic potential used in the Monte Carlo simulation, which also coincides with the above zero-temperature result. Meanwhile, the $-P$ case always give a uniform charge distribution between two layers (Fig.~\ref{mc}(b)).

\section{Details of the DFT calculations}

The density functional theory (DFT) calculations were here performed based on the
projected augmented wave (PAW) pseudopotentials using
the Vienna \textit{ab initio} simulation
package (VASP).\cite{Blochl:Prb,Kresse:Prb,Kresse:Prb96}
The valence states include the orbitals
$5s5p6s$, $3p4s3d$, $5d4f6s$, $4s4p5s$, $3p4s3d$ and $2s2p$ for Ba,
Ti, La, Sr, Mn and O, respectively. The electron-electron interaction
is described using the generalized gradient approximation (GGA) method.
The energy cutoff is $500$ eV and the $\varGamma$-center $k$-mesh
is $5\times5\times1$ for the superlattice.

According to the model study, an isotropic superexchange is needed,
with a weak in-plane coupling and a strong out-of-plane coupling.
This effect can be fulfilled by using an in-plane tensile strain
for the manganite bilayer.
In the present study, cubic KTaO$_3$ is chosen as the substrate with a
lattice constant $3.989$ \AA{}.
In addition, this substrate can also give a compressive
strain to BaTiO$_3$, which can enhance its ferroelectric polarization.

First, the atomic positions and lattice constants along the $c$-axis
were relaxed for pure La$_{0.75}$Sr$_{0.25}$MnO$_3$ and BaTiO$_3$
(with spontaneous ferroelectric polarization) respectively, by fixing
the in-plane crystal lattice constants as $3.989\times\sqrt{2}$ \AA{}
to match the KTaO$_3$ substrate.

Then, a superlattice structure consisting of $4$ layers of BaTiO$_3$
and a La$_{0.75}$Sr$_{0.25}$MnO$_3$ bilayer (for a total of $60$ atoms)
is stacked along the [001] axis, according to their individual
structures. Due to the doping of Sr, there are two types of
interfaces: (1)...-TiO$_2$-LaO-MnO$_2$-La$_{0.5}$Sr$_{0.5}$O-MnO$_2$-BaO-... and
(2)...-TiO$_2$-La$_{0.5}$Sr$_{0.5}$O-MnO$_2$-LaO-MnO$_2$-BaO-... .
The first one has a stronger polar interface, which will be studied
in the present work.

To simulate a robust $\pm P$ in such a thin ferroelectric layer with
asymmetric polar interfaces, the atomic positions for BaTiO$_3$ are frozen
during the atomic relaxation. Otherwise the $+P$ state may be depolarized
by the polar interfaces. In practice, this issue can be solved by using
a little thicker ferroelectric layers or poling it using electric fields.
In fact, with good ferroelectric materials (e.g. PbZr$_{1-y}$Ti$_y$O$_3$), $\pm P$ can be both stabilized even
when its thickness is only $3$ nm when attached to LSMO.\cite{Jiang:Nl}
The atoms of La$_{0.75}$Sr$_{0.25}$MnO$_3$ and the interfacial linking
oxygens are relaxed under different magnetic orders (ferromagnetic and
A-type antiferromagnetic) and $\pm P$, as the total energies converge.
Then, the total energies and magnetization are calculated for the relaxed structures.

\begin{figure}
\centering
\includegraphics[width=0.5\textwidth]{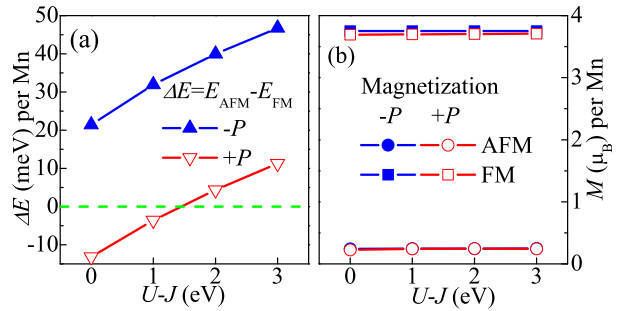}
\caption{(a) The energy difference per Mn between the FM
and AF states ($\Delta E=E_{\rm AF}-E_{\rm FM}$) as a function of
the effective $U_{\rm eff}$ for the $\pm P$ cases. If $\Delta E$ is
positive (negative), then the FM (AF) state is the ground state in our calculations.
(b) The total magnetization per Mn calculated as a function of $U_{\rm eff}$.
For the FM state, the value is close to saturation $3.75$ $\mu_B$ suggesting a robust
ferromagnetism. For the AF state, the (absolute value of) net magnetic moment
is $0.22-0.24$ $\mu_B$. Both these two values are almost independent of $U_{\rm eff}$
and $\pm P$.}
\label{U}
\end{figure}

In addition, the GGA+$U$ method in the Dudarev approach\cite{Dudarev:Prb}
was also tested by adopting the above relaxed lattices. The Hubbard $U$ was
added to the Mn's $3d$ orbitals. As shown in Fig.~\ref{U}(a), the switching
between the FM and AF states can occur in the low-$U$ region ($U_{\rm eff}=U-J$,
$U_{\rm eff}<\sim1.5$ eV), but it becomes ``un-switchable'' in the large-$U$ region
since the FM configurations dominate for both polarization directions.
In other words, with increasing $U$'s the system develops a stronger
FM tendency, in agreement with previous DFT calculations.\cite{Luo:Prl,Chen:Prb12}

In spite of this potential ``problem", the energy difference
between the FM and AF states in the $\sim40$ meV/per Mn scale suggests that the $\pm P$ switching could still
occur in practice by simply fine tuning other parameters in the experiments.
For example, one can use LSMO with a little higher doping $x$ chosen to be closer
to the AF state according to Ref.~\onlinecite{Chen:Prb12}. In our current DFT calculation,
we can only deal with the $x=1/4$ case due to cluster size limitations.
But in real experiments, considering the many magnetic phases of manganites and rich phase
diagrams, it is certainly
possible to find a proper material and proper substrates to realize the giant switch effect
proposed here. In other words, by no means our ideas are restricted to $x=1/4$.

In addition, some previous DFT studies found that the GGA method is better
than GGA+$U$ for doped metallic manganites,\cite{Luo:Prl,Burton:Prb} and
a weak $U$ ($\sim1-2$ eV) was found to be better than a large $U$.\cite{Chen:Prb12}
Then, the GGA+$U$ results suggesting that at large $U$ the switch does not occur
may need revision as well.

As shown in Fig.~\ref{U}(b), the total magnetizations of the FM and AF states
are almost unaffected by $U$, implying that the giant modulation of the
magnetization is quite stable if it can be realized in experiments.

The spin-orbit coupling, which is weak and not important in this case,
is not included in the DFT calculations.

\section{The ``dead layer'' problem of manganite ultra-thin films}

In ultra-thin manganite films a practical problem is caused by
the presence of the so-called ``dead layers''. When the thickness
of a LSMO thin film with composition $x\sim 0.3$ is reduced
to just a few unit cells (u.c.) the film becomes
non-magnetic and insulating. This is surprising because at this composition
LSMO is FM and metallic when in bulk form.
For example, in 2008 Huijben \textit{et al.}  \cite{Huijben:Prb}
and Tebano \textit{et al.} \cite{Tebano:Prl} it has been
reported that for LSMO thin films
grown on SrTiO$_3$ (STO) with a thickness below $8$ u.c., the metallicity
disappeared. However, despite this reported insulating behavior, thin
films of $5-6$ u.c.
remained FM although with a $T_{\rm C}$ reduced from $\sim300$ K
to $\sim100-150$ K and with a saturated magnetization reduction
to $1/3-1/4$ of the expected values.\cite{Huijben:Prb,Tebano:Prl}

It is very important to understand and overcome this ``dead layer''
problem considering the potential applications of LSMO films in
spintronics. For the results reported in the present publication this
effect is particularly important since our main conclusions have been obtained
using bilayers of LSMO, that may potentially contain dead layers.
However, there is considerable controversy regarding
the accurate value of the critical thickness and its underlying
mechanism. The critical thickness has been found to depend strongly
on several experimental
technical aspects, such as the oxygen partial pressures used
during the growth, the type of substrates,
annealing processes employed, and even on laser spots in
the PLD growth. Thus, the critical thickness varies
from group to group and even within the same group.

From the theoretical perspective, there are no intrinsic reasons
why LSMO thin films grown on a STO substrate must be ``dead''
if the crystal structure is without any defects. Both model Hamiltonian
and DFT calculations have reported a robust FM state for a few LSMO layers grown
on STO, e.g. as in our current work and Ref.~\onlinecite{Wang:Apl13}. In this sense, the underlying mechanism for the presence of
dead layers is likely related with imperfections in the LSMO ultra thin-films
such as atomic intermixing.

Although in Refs.~\onlinecite{Huijben:Prb,Tebano:Prl}, the orbital reconstruction
was associated with the existence of dead layers,
recent experiments attributed
an even larger contribution to local non-stoichiometric effects.
For example, in Ref.~\onlinecite{Peng:Arx}, compelling evidence was showed
that the intrinsic oxygen vacancy formation was the reason
for the presence of dead layers. And in Refs.~\onlinecite{Song:Sia,Li:AFM},
the layer-dependent non-stoichiometry was found to be affected
by details of the interface and surface.

As discussed in Ref.~\onlinecite{Peng:Arx}, this local non-stoichiometry may
be caused by the internal electrostatic field induced by the polar
discontinuity effect between the LSMO/substrate and LSMO/vacuum interfaces.
Thus, with the continuous
advances in the experimental techniques and with a better understanding
of the underlying mechanism, it is reasonable to expect that the dead layer
effect will be further suppressed in the future.

The polar discontinuity related with the LSMO/vacuum interface
can be easily eliminated
in superlattices with repeated interfaces, as used in our model system.
In fact, recent experiments reported that the intrinsic critical
thickness of the dead layers has an {\it upper} bound of
just $2$ u.c. per interface
in LSMO/STO superlattices.\cite{Kourkoutis:Pnas} In fact,
$4-5$ u.c. LSMO thin films exhibited
clear ferromagnetism above room temperature and remained metallic
below the Curie temperature, even in the presence of some
interfacial interdiffusion.\cite{Gray:Prb,Kourkoutis:Pnas}
Although these two experiments did not study even thinner cases,
it is likely that $4-5$ u.c. is not itself the critical limit for dead layers
in LSMO/STO superlattices since its Curie temperature and magnetization
are robust. In other words, it is to be expected that the ferromagnetism
will survive in even thinner cases.

Also, the polar discontinuity related
with the LSMO/substrate interface can be
partially eliminated with modern technology. For example,
in Refs.~\onlinecite{Peng:Arx,Boschker:AFM}, the magnetism and
conductance of LSMO thin films were improved by careful
interfacial engineering, even in the presence of open surfaces.

Finally, it is important to remark that the polar electrostatic
driving force to form the aforementioned intrinsic defects is {\it weaker}
in the present bilayer case than in thicker ones.
This tendency has also been observed in our simulation, as shown in Fig.~\ref{zt}(c-d).
According to the polar catastrophe scenario,\cite{Nakagawa:Nm} the electrostatic potential
between the top and bottom layers increases with the thickness of the films.
In bilayers only two layers are neighbors and the interference
between these neighboring interfaces significantly suppresses the polar
electrostatic potential, which will be helpful to further eliminate
defects. Experimental evidence is that in Ref.~\onlinecite{Song:Sia} the
layer-resolved non-stoichiometry was weaker in the $3$ u.c. case than
in the $5$ u.c. one.
In addition, the polar electrostatic effect has already been taken into account in our bilayer simulations.

In summary, recent experimental progress has clearly indicated that
the ``dead layer'' problem of LSMO thin films is mainly related with
the quality of the interface. Thus, it is a {\it technical} problem.
This problem is weaker in superlattices,
as described in the present study. And the critical thickness of
the ``dead layers'' has been significantly reduced in recent
years and this progress is going on. In fact, in the past an analogous story developed in the context of
dead layers in ferroelectric thin films, which also puzzled researchers
for several years. This effect has been successfully solved due to
improvements in the preparation of the samples.\cite{Fong:Sci}

\bibliographystyle{apsrev4-1}
\bibliography{../ref}
\end{document}